\documentclass[11pt]{article}
\usepackage{moriond,epsfig}
\usepackage{graphics}
\usepackage{psfrag}

\def\be{\begin{equation}}
\def\ee{\end{equation}}
\def\bea{\begin{eqnarray}}
\def\eea{\end{eqnarray}}

\def\Neqfour{{\cal N}=4}
\def\spa#1.#2{\left\langle#1\,#2\right\rangle}
\def\spb#1.#2{\left[#1\,#2\right]}
\newcommand{\ssplit}[2]{\mathrm{split}(#1\to P^{#2})}
\def\Del#1.#2.#3{\Delta_{(1)}(#1,#2;#3)}

\begin{document}
\vspace*{4cm}
\title{Multi gluon collinear limits from MHV amplitudes}

\author{P.~ Marquard, T.G.~Birthwright}

\address{Institute of Institute of Particle Physics Phenomenology, Department of Physics,
University of Durham, Durham DH13LE, UK}

\maketitle\abstracts{
We consider the multi-collinear limit of multi-gluon QCD
  amplitudes at tree level. We use the MHV rules for constructing
  colour ordered tree amplitudes and the general collinear
  factorisation
  formula to derive timelike splitting functions
  that are valid for specific numbers of negative helicity
  gluons and an arbitrary number of positive helicity gluons (or vice versa).
}

\section{Introduction}
\begin{picture}(0,0)(0,0)
\put(230,450){IPPP/05/17, DCPT/05/34, hep-ph/0505264}
\end{picture}
This talk is based on our recent paper\cite{Birthwright:2005ak}, where
we exploit the MHV
formalism to examine the singularity structure of tree-level
amplitudes when many gluons are simultaneously collinear.

The interpretation of $\Neqfour$ supersymmetric Yang-Mills theory and
QCD as a topological string propagating in twistor
space~\cite{Witten1}, has inspired a new and powerful framework for
computing tree-level and one-loop scattering amplitudes in Yang-Mills
gauge theory.  Notably, two distinct formalisms have been developed
for calculations of scattering amplitudes in gauge theory -- the `MHV
rules' of Cachazo, Svr\v{c}ek and Witten (CSW) \cite{CSW1}, and the
`BCF recursion relations' of Britto, Cachazo, Feng and Witten
\cite{BCF4,BCFW}.

Understanding the infrared singular behaviour of multi-parton
amplitudes is a prerequisite for computing infrared-finite cross
sections at fixed order in perturbation theory.  In general, when one
or more final state particles are either soft or collinear, the
amplitudes factorise. The first factor in this product is an amplitude
depending on the remaining hard partons in the process (including any
hard partons constructed from an ensemble of unresolved partons). The
second factor contains all of the singularities due to the unresolved
particles.  One of the best known examples of this type of
factorisation is the limit of tree amplitudes when two particles are
collinear.  This factorisation is universal and can be generalised to
any number of loops~\cite{Kosower:allorderfact}.

A useful feature of the MHV rules is that it is not required to set
reference spinors $\eta_\alpha$ and $\eta_{\dot\alpha}$ to specific
values dictated by kinematics or other reasons.  In this way, on-shell
(gauge-invariant) amplitudes are derived for arbitrary $\eta$'s, i.e.
without fixing the gauge.  By starting from the appropriate colour
ordered amplitude and taking the collinear limit, the full amplitude
factorises into an MHV vertex multiplied by a multi-collinear
splitting function that depends on the helicities of the collinear
gluons.  Because the MHV vertex is a single factor, the collinear
splitting functions have a similar structure to MHV amplitudes.
Furthermore, the gauge or $\eta$-dependence of the splitting function
drops out.

One of the main points of our approach is that, in order to derive all
required splitting functions we do not need to know the full
amplitude. Out of the full set of MHV-diagrams contributing to the
full amplitude, only a subset will contribute to the multi-collinear
limit. This subset includes only those MHV-diagrams which contain an
internal propagator which goes on-shell in the multi-collinear limit.
In other words, the IR singularities in the MHV approach arise
entirely from internal propagators going on-shell. This observation is
specific to the MHV rules method and does not apply to the BCF
recursive approach. 

The basic building blocks of the MHV rules approach~\cite{CSW1} are
the colour-ordered $n$-point vertices which are connected by scalar
propagators. These MHV vertices are off-shell continuations of the
maximally helicity-violating (MHV) $n$-gluon scattering amplitudes of
Parke and Taylor~\cite{ParkeTaylor,BG}.  They contain precisely two
negative helicity gluons.  Written in terms of spinor inner products,
they are composed entirely of the holomorphic products $\spa{i}.{j}$
of the right-handed (undotted) spinors, rather than their
anti-holomorphic partners $\spb{i}.{j}$, \be
A_n(1^+,\ldots,p^-,\ldots,q^-,\ldots,n^+) = \frac{\spa{p}.{q}^4}{
  \spa1.2 \spa2.3 \cdots \spa{n-1,}.{n} \spa{n}.{1} },
\label{MHV}
\ee
where we introduce the common notation
$\spa{p_i}.{p_j}=\spa{i}.{j}$ and $\spb{p_i}.{p_j}=\spb{i}.{j}$.
By connecting MHV vertices,  amplitudes involving more
negative helicity gluons can be built up.

The factorisation properties of amplitudes in the infrared
play several roles in developing higher order
perturbative predictions for observable quantities. First, a detailed
knowledge of the structure of unresolved emission enables phase space
integrations to be organised such that the infrared singularities due to soft
or collinear emission can be analytically
extracted.    Second, they
enable large logarithmic corrections to be identified and resummed.
Third, the collinear limit plays a crucial role in the unitarity-based method
for loop calculations.
In general, to compute a cross section at N$^n$LO, one requires detailed
knowledge of the infrared factorisation functions describing the unresolved
configurations for $n$-particles at tree-level, $(n-1)$-particles at one-loop
etc.

\section{Examples}
In this section I'll give some simple examples of how to obtain the
splitting amplitudes starting from the general factorisation formula.
For more details and the full results for up to six gluons please see
ref. \cite{Birthwright:2005ak}.  The general factorisation formula
describes the behaviour of the amplitude in the collinear limit. When
$n$ particles become collinear the amplitude is given by
\begin{eqnarray}
  \label{eq:factorise}
  A_N(1^{\lambda_1},\ldots,N^{\lambda_N}) &\to&
  \ssplit{{1}^{\lambda_{1}},\ldots,n^{\lambda_n} }{\lambda}
  \times
  A_{N-n+1}((n+1)^{\lambda_{n+1}},\ldots ,N^{\lambda_N},P^\lambda).\nonumber 
\end{eqnarray}
where $\mathrm{split}$ denotes the splitting function. 
The simplest example is the case of $\ssplit{{1}^{+},\ldots,n^{+}
}{+}$ which can be obtained directly from a single MHV vertex namely
\begin{equation}
  A(1^+,\ldots,n^+,(n+1)^+,(n+2)^-,(n+3)^-)
  =\frac{\spa{n+2}.{n+3}^4}{\prod_{i=1}^{n+2} \spa{i}.{i+1}}.
\end{equation}
Now taking particles $1,\ldots,n$ collinear, assuming that 
\begin{equation}
  \sum_{i=1}^n p_i = P \quad \mbox{and} \quad p_i \to z_i P \quad
  \Rightarrow \quad  \spa{i}.{n+1} \to \sqrt{z_i}\spa{P}.{n+1}
\end{equation}
in the collinear limit, we obtain
\begin{eqnarray*}
  \lefteqn{A_{n+3}(1^+,\ldots,n^+,(n+1)^+,(n+2)^-,(n+3)^-)=} &&\\
&&\underbrace{\frac{\spa{n+2}.{n+3}^4}{\spa{P}.{n+1}\spa{n+1}.{n+2}\spa{n+2}.{n+3}\spa{n+3}.{P}}
}_{A_4((n+1)^+,(n+2)^-,(n+3)^-,P^+)} 
\times \underbrace{\frac{1}{\sqrt{z_1}\sqrt{z_n}\prod_{i=1}^{n-1}\spa{i}.{i+1}}}_{\ssplit{1^+,\ldots,n^+}{+}}
\end{eqnarray*}
where we can directly read off the splitting function.
Similarly we can obtain the splitting function
$\ssplit{1^+,\ldots,i^-,\ldots,n^+}{-}$.
\begin{equation}
  \ssplit{1^+,\ldots,i^-,\ldots,n^+}{-} = \frac{z_i^2}{\sqrt{z_1}\sqrt{z_n}\prod_{i=1}^{n-1}\spa{i}.{i+1}}
\end{equation}
by starting from the amplitude
$A(1^+,\ldots,i^-\ldots,n^+,(n+1)^+,(n+2)^+,(n+3)^-)$.

\begin{figure}[t!]
    \centering
    \caption{NMHV diagrams contributing to $\ssplit{1^+,\ldots,i^-,\ldots,n^+}{-} $.  Negative helicity gluons are indicated
    by solid lines, while
    arbitrary numbers of positive helicity gluons emitted from each vertex are shown as dotted arcs.}
\label{fig:m1}
    \psfrag{i+1+}{$~$}
    \psfrag{j+1+}{$~$}
    \psfrag{j+}{$j+$}
    \psfrag{i+}{{\small $i+$}}
    \psfrag{m1-}{$m_1^-$}
    \psfrag{m2-}{$m_2^-$}
    \begin{center}
        \includegraphics[width=5cm]{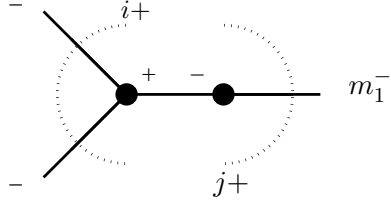}
    \end{center}
\end{figure}

A more complicated example is the splitting function
$\ssplit{1^+,\ldots,i^-,\ldots,n^+}{+} $. This splitting amplitude can
not be obtained from MHV amplitudes alone. In this case we have to
consider NMHV amplitudes which yields to more complicated expressions.
In general the difficulty of the problem increases with $\Delta M$,
which is given by the change of the number of gluons with negative
helicities. The first two examples were of the type $\Delta M = 0$,
while this example is of the type $\Delta M =1$. In this case only the
diagram in figure \ref{fig:m1} contributes since it is the only
diagram in which the propagator becomes onshell in the collinear
limit.
Starting with the MHV representation of this diagram and performing
the same replacements as above we obtain
\begin{equation}
\label{mnpp}
  \ssplit{1^+,\ldots,i^-,\ldots,n^+}{+} 
= \frac{1}{\sqrt{z_1 z_n} \prod_{l=1}^{n-1} \spa{l,}.{l+1} }\bigg(
  \sum_{i=0}^{m_1-1} \sum_{j=m_1}^{n}
  \frac{\Del{i}.{j}.{m_1}^4}{D(i,j,q_{i+1,j})} \bigg)\ ,
\end{equation}
where we define
\be
\label{D}
D(i,j,q)=\frac{q_{i+1,j}^2}{\spa{i,}.{i+1} \spa{j,}.{j+1}}
\Del{i}.{j}.{i}\Del{i}.{j}.{i+1}\Del{i}.{j}.{j}\Del{i}.{j}.{j+1}\ .
\ee
and
\begin{equation}
  \label{eq:7}
  \Del{i}.{j}.{a} = \sum_{l=i+1}^{j}\spa{a}.{l}\sqrt{z_l} .
\end{equation}
Specific results can immediately be obtained from this expression, e.g.
\bea
\label{eq:triple1}
\ssplit{1^-, 2^+, 3^+}{+} &=& {\frac {\spa{1}.{2}{z_{{2}}}^{2}}
{\sqrt {z_{{1}}z_{{2}}z_{{3}}}s_{{1,2}}
 \left( z_{{1}}+z_{{2}} \right)  \left( \spa{1}.{3}\sqrt {z_{{1}}}+\spa{2}.{3}\sqrt {z_{{2}}}
 \right) }}
 \nonumber\\
&+&
{\frac { \left( \spa{1}.{2}\sqrt {z_{{2}}}+\spa{1}.{3}\sqrt {z_{{3}}} \right) ^{3}}
 {s_{{1,3}} \spa{1}.{2}\spa{2}.{3} \left( \spa{1}.{3}\sqrt {z_{{1}}}+\spa{2}.{3}\sqrt {z_{{2}}}
  \right) }}  
\eea

\section{Conclusion}
In \cite{Birthwright:2005ak} we have considered the collinear limit of
multi-gluon QCD amplitudes at tree level. We have used the new MHV
rules for constructing colour ordered amplitudes from MHV vertices
together with the general collinear factorisation formula to derive
timelike splitting functions that are valid for specific numbers of
negative helicity gluons with an arbitrary number of positive helicity
gluons (or vice versa). In this limit, the full amplitude factorises
into an MHV vertex multiplied by a multi-collinear splitting function
that depends on the helicities of the collinear gluons. These
splitting functions are derived directly using MHV rules. Out of the
full set of MHV-diagrams contributing to the full amplitude, only the
subset of MHV-diagrams which contain an internal propagator which goes
on-shell in the multi-collinear limit contribute.

We find that the splitting functions can be characterised by $\Delta
M$, the difference between the number of negative helicity gluons
before taking the collinear limit, and the number after.  $\Delta M+1$
also coincides with the number of MHV vertices involved in the
splitting functions.  Our main results are splitting functions for
arbitrary numbers of gluons where $\Delta M =0,1,2$. Splitting
functions where the difference in the number of positive helicity
gluons $\Delta P = 0,1,2$ are obtained by the parity transformation.
These general results are sufficient to describe {\em all} collinear
limits with up to six gluons.  We have given explicit results for up
to four collinear gluons for all independent helicity combinations,
which numerically agree with the results of Ref.~\cite{delduca},
together with new results for five and six collinear gluons. This
method can be applied to higher numbers of negative helicity gluons,
and via the MHV-rules for quark vertices, to the collinear limits of
quarks and gluons \cite{Birthwright:2005vi}.

\section*{Acknowledgments}
TGB acknowledges the award of a PPARC studentship.

\end{document}